\documentclass[twocolumn, tradiabstract]{aa} % for a paper on 1 column

\def\rmit#1{{\it #1}}              %% italics (RR mode, Kluwer)
\def\specchar#1{{\sc #1}}

\def\ie{\rmit{i.e.}}

\def\degree{\hbox{$^\circ$}}
\def\arcsec{\hbox{$^{\prime\prime}$}}
\def\FeI{\mbox{Fe\,\specchar{i}}}
\def\NiI{\mbox{Ni\,\specchar{i}}}

\def\BaII{\mbox{Ba\,\specchar{ii}}}
\def\CaIIK{\mbox{Ca\,\specchar{ii}\,\,K}}       %% use \CaIIK\ for space
\def\CaIIH{\mbox{Ca\,\specchar{ii}\,\,H}}       %% use \CaIIH\ for space

\def\pun{\stackrel{}{\mbox{.}}}
\def\farcs{$\stackrel{\prime\prime}{\pun}$}

\usepackage{natbib}
\usepackage{graphicx}
\begin{document}
   \title{On the origin of facular brightness}

\author{R. Kostik\inst{1}, E. Khomenko\inst{2,3,1}}

\institute{Main Astronomical Observatory, NAS, 03680, Kyiv, Ukraine\\ \email{kostik@mao.kiev.ua} \and Instituto de Astrof\'{\i}sica de Canarias, 38205 La Laguna, Tenerife, Spain
\and Departamento de Astrof\'{\i}sica, Universidad de La Laguna, 38205, La Laguna, Tenerife, Spain\\
}

\date{Received XXX, 2015; accepted xxx, 2015}

\abstract{This paper studies the dependence of the \CaIIH\ line core brightness on the strength and inclination of photospheric magnetic field, and on the parameters of convective and wave motions in a facular region at the solar disc center. We use three simultaneous datasets obtained at the German Vacuum Tower Telescope (Observatorio del Teide, Tenerife): (1) spectra of \BaII\ 4554 \AA\ line registered with the instrument TESOS to measure the variations of intensity and velocity through the photosphere up to the temperature minimum; (2) spectropolarimetric data in \FeI\ 1.56 $\mu$m lines (registered with the instrument TIP II) to measure photospheric magnetic fields; (3) filtergrams in \CaIIH\ that give information about brightness fluctuations in the chromosphere. The results show that the \CaIIH\ brightness in the facula strongly depends on the power of waves with periods in the 5-min range, that propagate upwards, and also on the phase shift between velocity oscillations at the bottom photosphere and around the temperature minimum height, measured from \BaII\ line. The \CaIIH\ brightness is maximum at locations where the phase shift between temperature and velocity oscillations lies within 0\degree-100\degree. There is an indirect influence of convective motions on the \CaIIH\ brightness. Namely, the higher is the amplitude of convective velocities and the larger is the height where they change their direction of motion, the brighter is the facula. Altogether, our results lead to conclusions that facular regions appear bright not only because of the Wilson depression in magnetic structures, but also due to real heating.}


\maketitle

%________________________________________________________________
\section{Introduction}

High spatial resolution observations reveal that facular regions break into clusters of bright points, small pores, and facular granular cells \citep{Dunn+Zirker1973, Title+etal1992, Berger+etal2004, Lites+etal2004, Narayan+Scharmer2010, Kobel+etal2011, Viticchie+etal2011}. It is believed that these features appear as a consequence of the presence of small scale magnetic elements (or flux tubes) of the size of hundreds of kilometers and magnetic field strength of the order of 1-2 kG  \citep{Stenflo1973, Solanki1993}. 
According to numerous theoretical and empirical models \citep{Spruit1976, Knoelker+etal1988, Grossmann-Doerth+etal1994, Topka+etal1997, Okunev+Kneer2005, Steiner2005}, the walls of these tubes are hot, and the temperature at the bottom depends on the diameter of the tube. The bottom is cold if the diameter $d$ is greater than 300 km, and it is hot if $d <300$ km. Because of the magnetic pressure, the temperature in a thick tube is lower than at the surrounding atmosphere at the same geometrical height. Therefore the bottom of the tube becomes dark in observations. But if the tube is sufficiently narrow, it can be heated by horizontal radiative transfer and becomes visibly bright. The contrast of the tube depends not only on its diameter, but also on the strength of the magnetic field of the tube, in a way that the larger is the field, the darker is the bottom of the tube. 
This latter dependence makes it possible to test theoretical models by means of observations. 

%%%%%%%%%%%%%%%%%%%%%%%%%%%%%%%%%%%%%%%%%%%%%%%%%%%%%%%%%%%%%%
\begin{figure*}
\centering
\includegraphics[width=5.5cm]{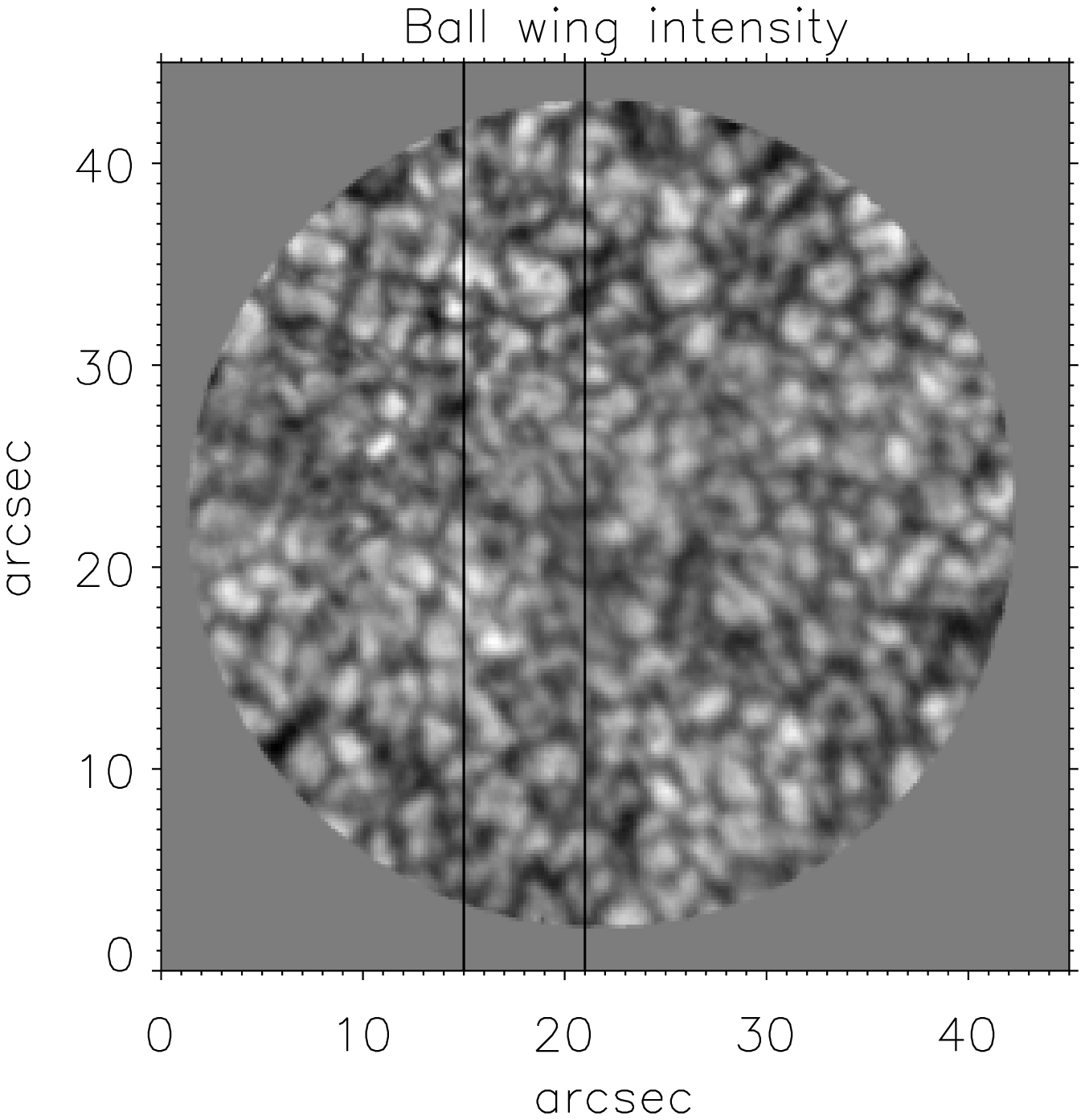}
\includegraphics[width=5.5cm]{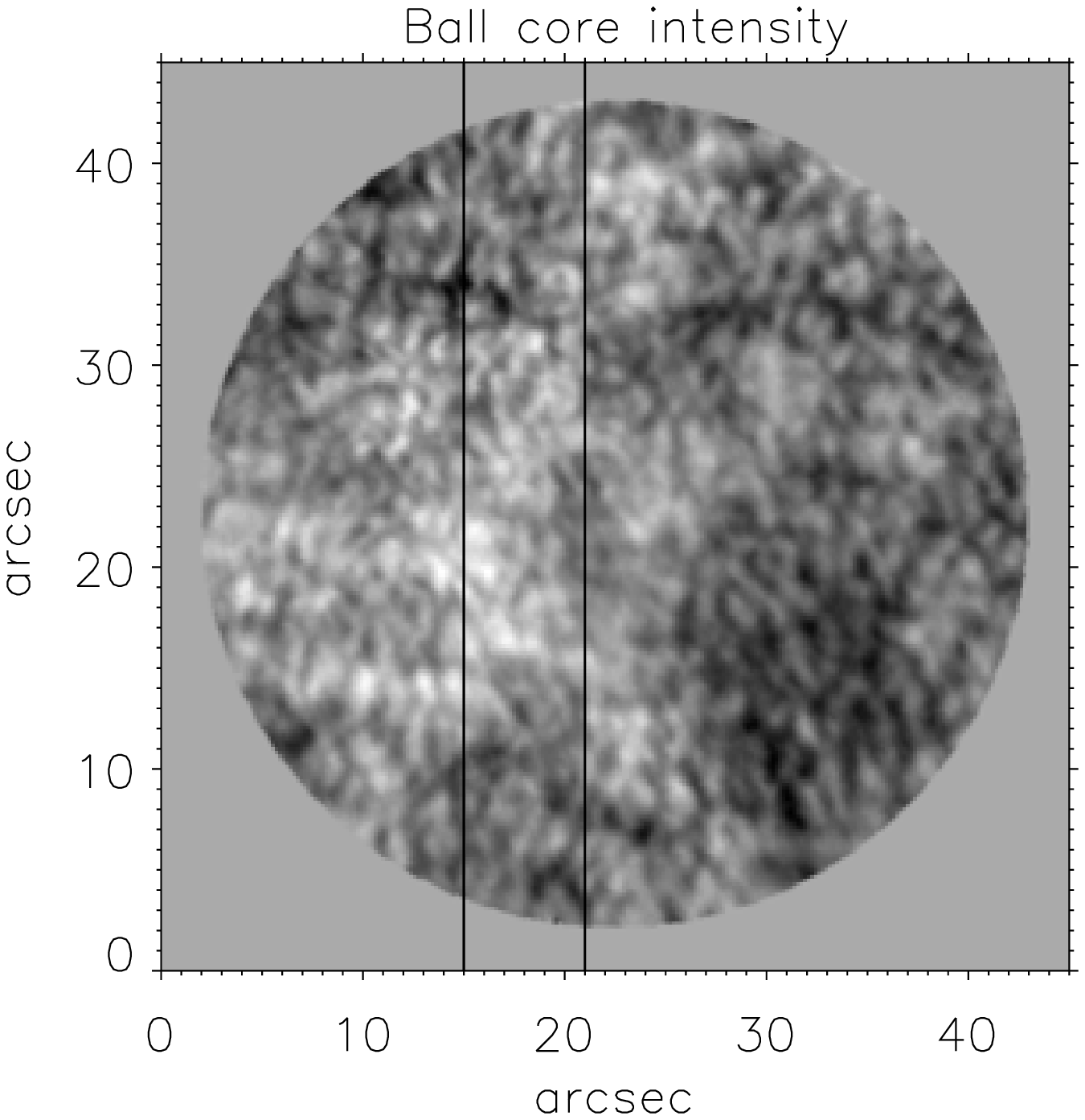}
\includegraphics[width=6.5cm]{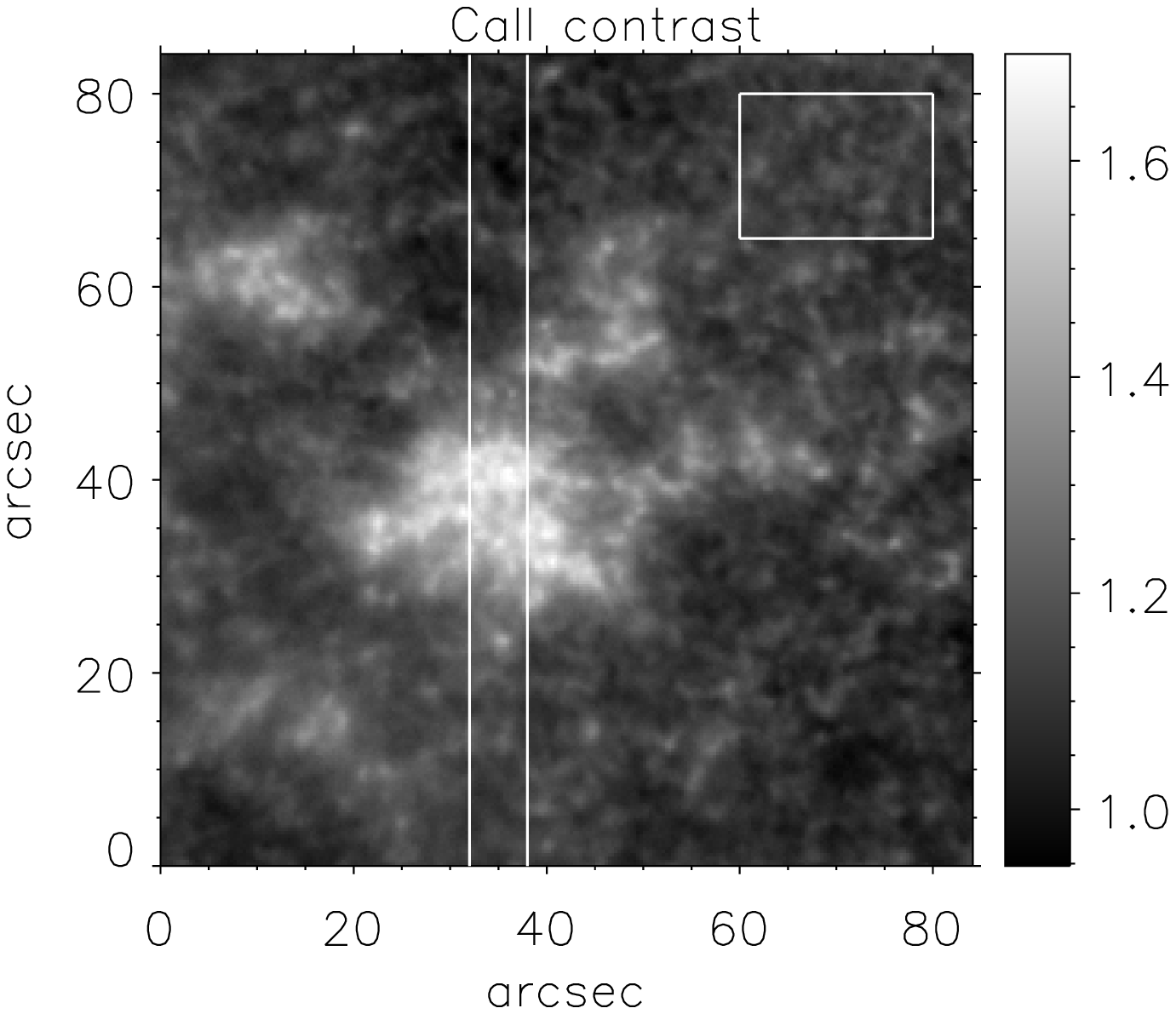}
\caption{Images of the complete field of view as detected in \BaII\ line wing (left), line core (center) and a \CaIIH\ contrast. The vertical lines mark the area scanned by TIP and used for the scientific analysis in this work. The rectangle area in the right panel  indicates the quiet area used as a reference for the definition of \CaIIH\ contrast.} \label{fig:fov}
\end{figure*}
%%%%%%%%%%%%%%%%%%%%%%%%%%%%%%%%%%%%%%%%%%%%%%%%%%%%%%%%%%%%%%

Indeed, observations of the solar disc center given in \citet{Frazier1971, Title+etal1992, Topka+etal1992, Montagne+etal1996, Topka+etal1997, Kobel+etal2011, Kostyk2013} show that the absolute value of continuum contrast of facular regions (without considering separately granules and intergranular lanes) depends on the magnetic field, though the observational findings are somewhat contradictory. \citet{Frazier1971} shows that the continuum contrast of the facula at the solar disc center and the magnetogram signal are correlated in a complicated manner. Below 200 G the contrast increases with the magnetograph signal, while after this value the continuum becomes monotonically darker until the pore and sunspot signals are reached. The \CaIIK\ emission also increases till about 500 G. \citet{Title+etal1992} find that the line center brightness of the \NiI\ 6768 \AA\ line is enhanced in the plage compared to the quiet area with a linear dependence between the magnetogram and the brightness, till about 600 G, and is decreased afterwards due to the presence of pores. The continuum contrast is unaffected by the plage magnetic flux till about the same strength, and then decreases. \citet{Topka+etal1992} present evidences that the continuum contrast at 5000 \AA\ of the facula close to the disc center is about 3\% less, which is contrary to many earlier observations. Outside of the disc center, the facular becomes brighter in continuum after about 20\degree. In their latter work, \citet{Topka+etal1997} show the decrease of the continuum contrast with the strength of the magnetogram signal at the disc center, and explain it in terms of the model of a flux tube with hot walls and cool floor.  In the later work by \citet{Montagne+etal1996} it was shown that the continuum brightness increase is associated with the presence of small scale magnetic elements in intergranular lanes, leading to an increase of brightness from 0 to 400 G, and a decrease afterwards, associated to the presence of larger structures. The authors used spectra of the \FeI\ 6301 and 6302 \AA\ lines and the entire profile was used to calculated the dependencies which is should give more precise results than in  \citet{Title+etal1992} and \citet{Topka+etal1992}  where the combination of non-simultaneous observations in the two wings of a spectral line was used. In a newer observations by Hinode, with much higher spatial resolution and when the magnetic field strength (not flux) was obtained using inversions,  \citet{Kobel+etal2011} finds that relationship between contrast and apparent longitudinal field strength exhibits a peak at around 700 G both for the quiet Sun network and active region plage, while earlier studies only found a monotonic decrease in active regions. The contrast possibly depends on the size of magnetic elements and not on their strength (the intrinsic strength is more or less constant in the kG range). \citet{Kostik+Khomenko2012} claim that the observed dependence between brightness and contrast is a consequence of the different distributions of magnetic field strength in granules and inter granular lanes. The histogram of the magnetic field in the facula has two maxima, one at the hecto-G and on a at the kG value. The first maximum has to do with granules, where the field is generally weaker, while the second one is related to intergranular lanes. If one considers the dependence between the contrast and the magnetic field strength selecting only intergranular lanes, then it turns out that the contrast remains practically unchanged with magnetic field increasing from 400 to 1600 G, see Figure 4 in \citet{Kostik+Khomenko2012}. Given all above, the observational dependence between the brightness and other quantities related to granulation needs further studies in order to clarify if solar faculae are indeed conglomerates of magnetic flux tubes. 
 
The purpose of this work is to deepen the study of the contrast of facular regions at the solar disc center, as observed at the core of  \CaIIH\ 3968 \AA\ line. Below we provide the dependences between the  \CaIIH\  contrast and the strength and inclination of the magnetic field, as well as various other parameters of solar granular and wave motions. 

%%%%%%%%%%%%%%%%%%%%%%%%%%%%%%%%%%%%%%%%%%%%%%%%%%%%%%%%%%%%%%
\begin{figure}[!th]
\centering
\includegraphics[width=8.0cm]{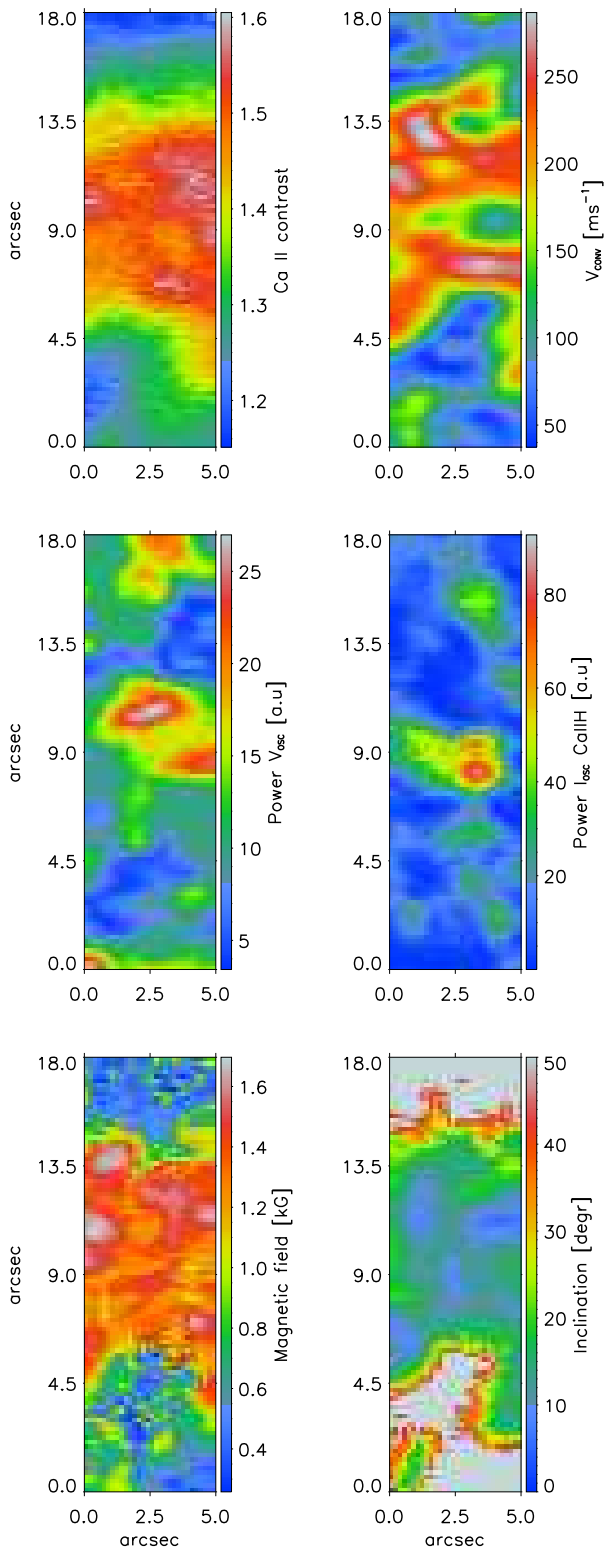}
\caption{Upper row: maps of \CaIIH\ contrast (left)  and time-average amplitude of the convective velocity at the bottom photosphere (right). Middle row: power of oscillatory velocity component in the upper photosphere (left) and power of oscillatory component of \CaIIH\ intensity (right). Bottom row: map of the magnetic field strength (left) and inclination with respect to the normal to the surface (right) in the observed field of view obtained after inversions of \FeI\ lines.} \label{fig:maps}
\end{figure}
%%%%%%%%%%%%%%%%%%%%%%%%%%%%%%%%%%%%%%%%%%%%%%%%%%%%%%%%%%%%%%

%%%%%%%%%%%%%%%%%%%%%%%%%%%%%%%%%%%%%%%%%%%%%%%%%%%%%%%%%%%%%%
\begin{figure}[!ht]
\centering
\includegraphics[width=8cm]{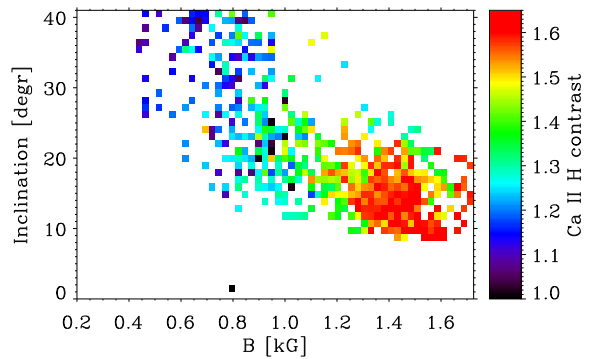}
\caption{Bi-dimensional representation of the \CaIIH\ contrast as a function of photospheric magnetic field strength and inclination. Each color rectangle in the scatter plot indicates the mean value of the contrast for a given range of field strengths and inclinations. The white area are the locations where particular combination of inclination and field strength is not present in our data.} \label{fig:caii_mf_2d}
\end{figure}
%%%%%%%%%%%%%%%%%%%%%%%%%%%%%%%%%%%%%%%%%%%%%%%%%%%%%%%%%%%%%%

\section{Observations and data reduction}

Following our previous papers from this series \citep{Kostik+Khomenko2012, Kostik+Khomenko2013}, we used observations performed at the German Vacuum Tower Telescope (VTT) located at the Observatorio del Teide in Iza\~na, Tenerife. The dataset was taken on  13th of November 2007. Three wavelength regions were observed simultaneously: \FeI\ $\lambda15643-15658$ \AA\ using TIP-II \citep{Collados2007}, \BaII\ $\lambda4554$ \AA\ using TESOS \citep{Tritschler+etal2002}, and \CaIIH\ $\lambda3968$ \AA\ using a broad-band filter at the VTT. A facular area close to the solar disc center at S05E04 was selected using filtergrams in \CaIIH\ line. Figure \ref{fig:fov} gives an overview of the observed area in TESOS (left and middle images) and a filtergram in \CaIIH. The vertical lines mark the area that was scanned by TIP. This figure allows to appreciate the quality of the observational material. It shows that no strong activity was present in the observed area. The granulation appress almost  undisturbed, no pores are present. The line core images in \BaII\ (middle panel) show suspicions brightness at the same locations as in \CaIIH\ filtergram that correspond to the facular area under study. The data treatment is described in detail in  \citet{Kostik+Khomenko2012, Kostik+Khomenko2013}. In the current work we study temporal variations in an area of a size of 5\farcs5$\times$18\farcs5 during 34 min and 41 sec of observations. The width of the spectrograph slit was of 0\farcs35. Our dataset contains: 
\begin{itemize}
\item Five TIP scan of Stokes spectra of \FeI\ lines at 1.56 $\mu$m, repeated every 6 min 50 sec  with a pixel size of 0\farcs185 and a spectral sampling of 14.73 m\AA/px. 
\item Time series of \BaII\ 4554 \AA\ monochromatic images with a temporal resolution of 25.6 sec, and pixel size of  0\farcs089,  tuned along the spectral line with spectral sampling of 16 m\AA/px between successive images. 
\item Time series of \CaIIH\ filtergrams with a temporal cadence of 4.93 sec, and a pixel size of 0\farcs123.
\end{itemize}
The seeing-limited angular resolution at the time of observations was no more than 0\farcs5-1\arcsec\ in the blue range of the spectrum at 4500 \AA, and around 2\arcsec\ in the infrared.

We obtained maps of the magnetic field strength and inclination by inverting the Stokes parameters of the \FeI\ 15648 and 15652 \AA\ lines  using the SIR inversion code \citep{RuizCobo+delToroIniesta1992}, see the details in \citet{Kostik+Khomenko2013}. Both the field strength and the inclination were assumed to be constant with height. The intensity and velocity oscillations were found using the $\lambda$-meter technique \citep{Stebbins+Goode1987} at 14 levels along the \BaII\ line profile \citep[\ie\ 14 heights in the photosphere, see][]{Shchukina+etal2009}. They are defined as follows \citep{Kostik+Khomenko2012}:
\begin{eqnarray}
\delta I(t,x,W) & = & I(t,x,W) - \bar{I}(W)  \\
\nonumber
\delta V_r(t,x,W) & = & V_r(t,x,W) - \bar{V_r}(W)  \\
\nonumber
\delta V_b(t,x,W) & = & V_b(t,x,W) - \bar{V_b}(W)
\end{eqnarray}
where $\delta V_r$ and $\delta V_b$ are red and blue wing velocities. In the equations above,  $W$ means one of 14 reference widths along the line profile, $x$ is spatial position and $t$ is time. The average intensity levels, $\bar{I}(W)$, and  blue and red wing reference positions $\bar{V_r}(W)$ and $\bar{V_b}(W)$ were obtained from the spatially and temporally averaged \BaII\ profile. The final velocity fluctuations are obtained from red and blue velocities as usual:
\begin{equation}
\delta V(t,x,W) =(\delta V_r(t,x,W) +\delta V_b(t,x,W))/2
\end{equation}

The convective and oscillatory components of the velocity and intensity variations were split using the $k-\omega$ diagram \citep{Khomenko+Kostik+Shchukina2001, Kostik+etal2009}. 

The \CaIIH\ contrast was calculated with respect to its average value of the quiet area of the same set of observations, $\bar{I}_{Ca}$, see the rectangular area indicated in the right panel of Figure \ref{fig:fov}:

\begin{equation} \label{eq:deltac}
\delta C(t,x)  =  I_{Ca}(t,x)/\bar{I}_{Ca} - 1
\end{equation}
where $I_{Ca}(t,x)$ is intensity in \CaIIH\ filtergram at a given point of a scan.

Therefore, at each observed pixel of the 5\farcs5$\times$18\farcs5 area, we have:
\begin{itemize}
\item The strength and inclination of the magnetic field at the deep photosphere.
\item Convective and oscillatory variations of the velocity and intensity measured along the photosphere (from 0 to 650 km) from the \BaII\ profile.
\item Chromospheric intensity variations from the core of \CaIIH\ line at a height of about 1000 km.
\end{itemize}

Figure \ref{fig:maps} gives an overview of the above variables in the observed field of view.  The comparison between the first two maps from the upper row of this figure with the magnetic field map shows that at locations with stronger magnetic field the \CaIIH\ contrast is enhanced. The amplitude of the convective velocity also shows larger values co-spatial with the areas with larger \CaIIH\ contrast, however its distribution is less homogeneous and more patchy. The power of oscillations of both velocity and \CaIIH\ intensity is large at the middle of the observed area, coinciding with the center of the magnetic area. The patch of enhanced oscillation power is smaller than the area of enhanced  \CaIIH\ contrast. The comparison between the maps of inclination and magnetic field strength  reveals that in the areas where the magnetic field is large (red area) the inclination varies between 0\degree\ and 30\degree. Outside of this patch the magnetic signal is weaker and we can not reliable measure the magnetic field inclination. In the analysis below we only use the area where the inclination does not exceed 40\degree. The correlations between all above quantities are discussed below and confirm the visual impression. 

Notice that the location of the observed region close to the disc center at S05E04 provides that the measured line-of-sight velocity corresponds to the vertical velocity. Therefore the velocity was suitably strong and could be measured reliably.  The inclination angles derived from inversions are good proxies for those with respect to the normal to the solar surface.

%%%%%%%%%%%%%%%%%%%%%%%%%%%%%%%%%%%%%%%%%%%%%%%%%%%%%%%%%%%%%%
\begin{figure*}
\centering
\includegraphics[width=12cm]{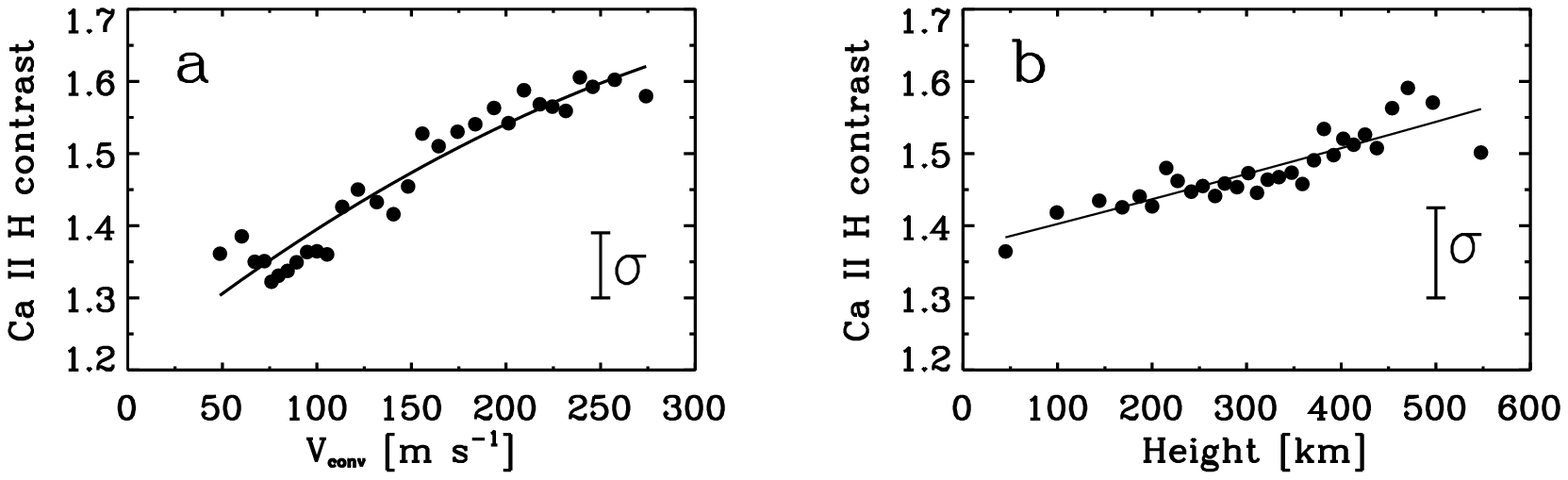}
\includegraphics[width=5cm]{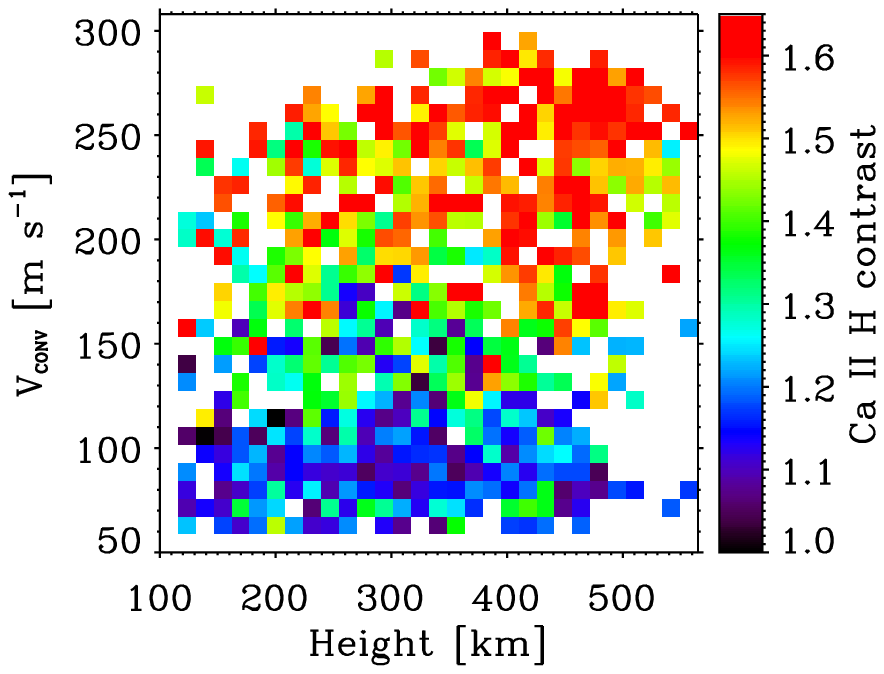}
\caption{\CaIIH\ contrast as a function of the parameters of convection. {\it Left panel:} contrast as a function of the amplitude of convective motions at the bottom photosphere. {\it Middle panel:} contrast as a function of height where convective velocities change their sign. Each point represents an average value over a bin containing an equal number of data points on the abscissa axis. The uncertainty $\sigma$ is measured as a standard deviation between the average and the individual values for each bin, and then averaged for all bins. {\it Right panel: } bi-dimensional representation of the \CaIIH\ contrast as a function parameters of convection. Each color rectangle in the scatter plot indicates the mean value of the contrast for a given range of convective velocity and height of sign reversal.  The white area are the locations where particular combination of convective velocities and reversal heights is not present in our data. } \label{fig:caii_conv}
\end{figure*}
%%%%%%%%%%%%%%%%%%%%%%%%%%%%%%%%%%%%%%%%%%%%%%%%%%%%%%%%%%%%%%

%%%%%%%%%%%%%%%%%%%%%%%%%%%%%%%%%%%%%%%%%%%%%%%%%%%%%%%%%%%%%%
\begin{figure*}
\centering
\includegraphics[width=12cm]{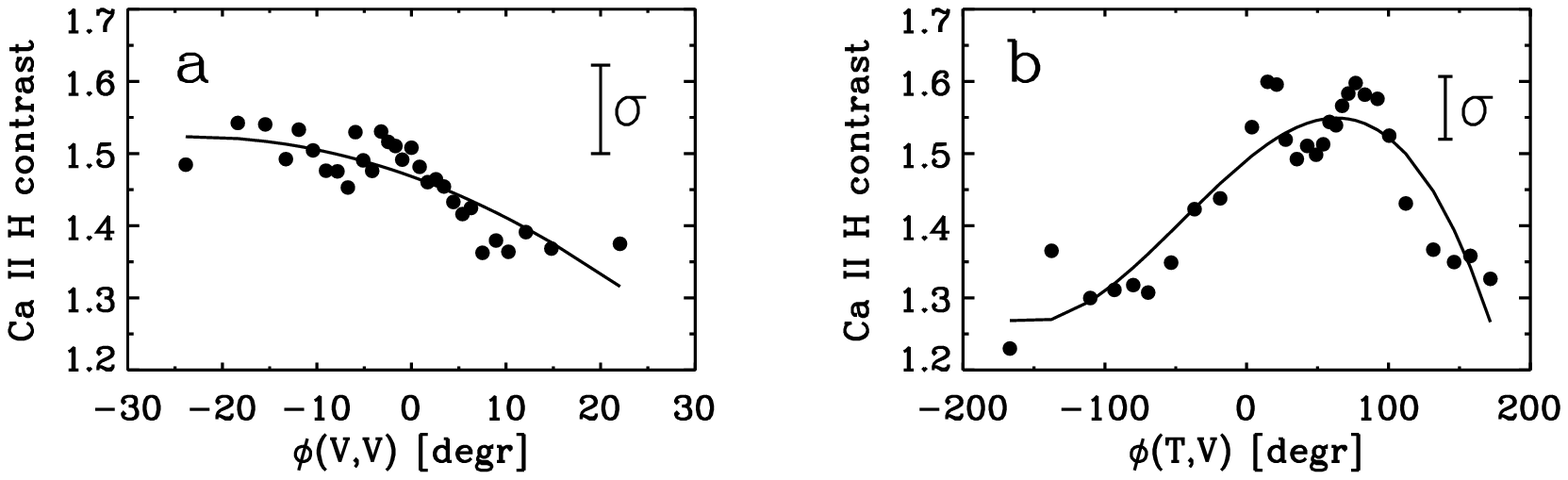}
\includegraphics[width=5cm]{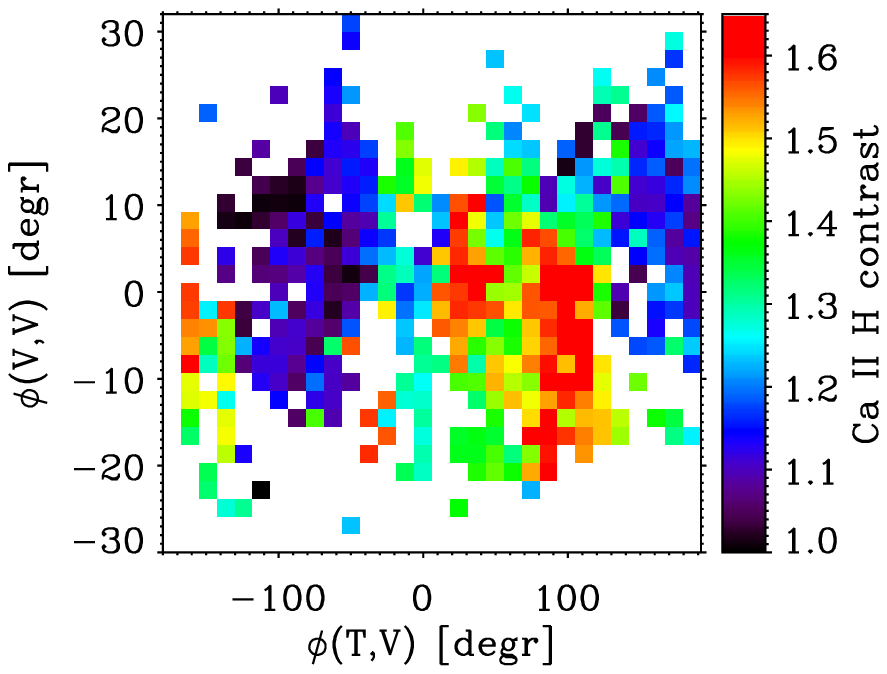}
\includegraphics[width=12cm]{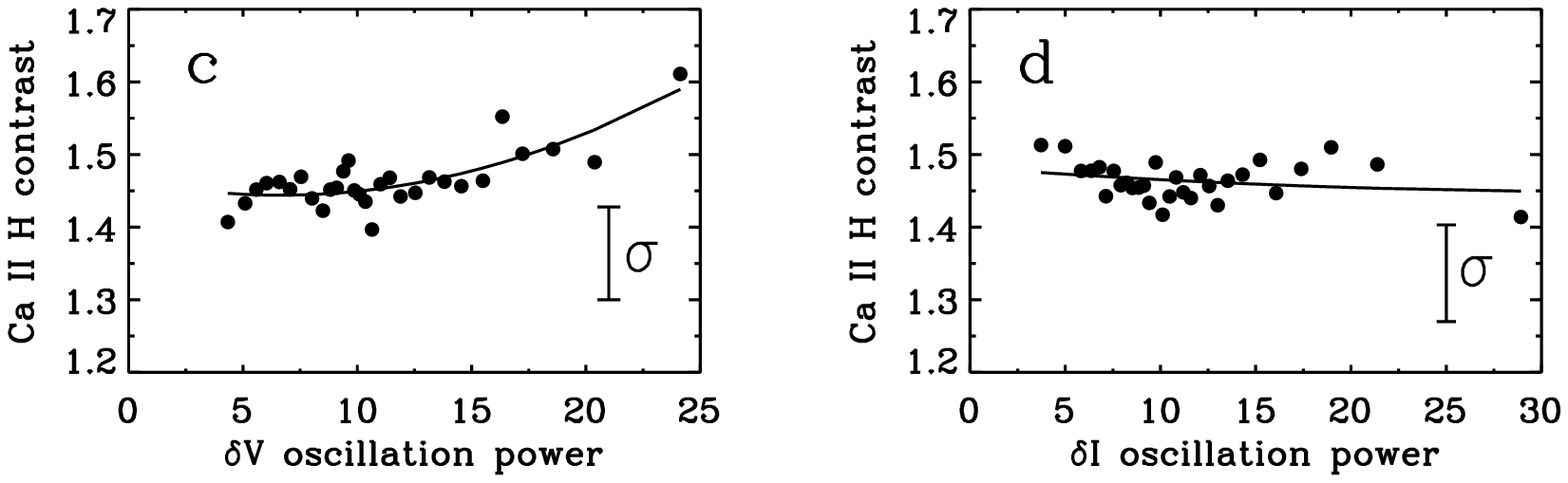}
\includegraphics[width=5cm]{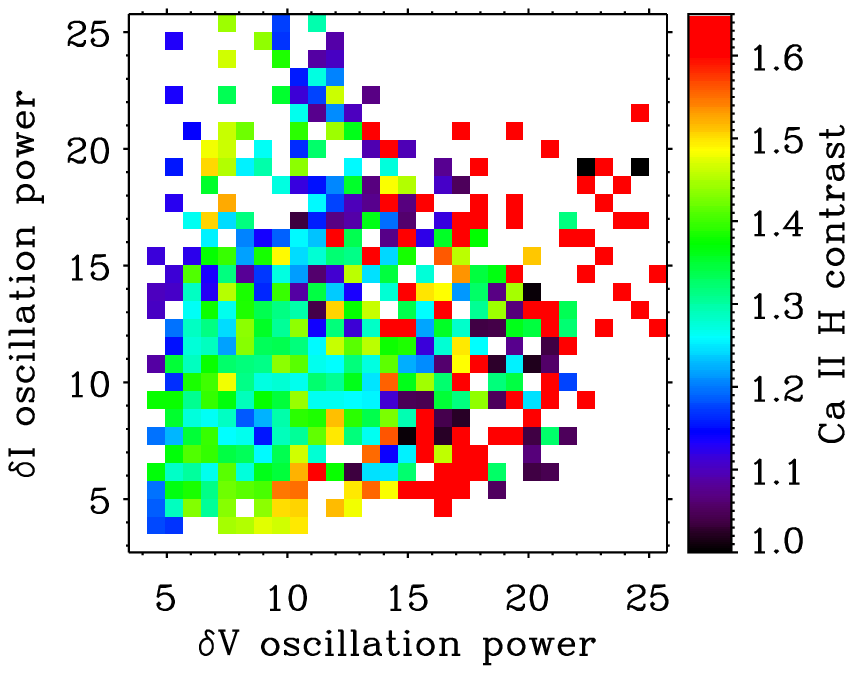}
\caption{\CaIIH\ contrast as a function of the parameters of wave motions. {\it Upper left panel:} contrast as a function of the phase shift between velocity oscillations at the bottom and upper photosphere for a period with maximum power, $\phi(V, V)$. {\it Upper middle panel:} contrast as a function of the phase shift between the temperature and velocity oscillations at the base of the photosphere, $\phi(T, V)$. {\it Upper right panel:} bi-dimensional representation of the \CaIIH\ contrast as a function of $\phi(V, V)$ and $\phi(T,V)$. {\it Bottom {\bf left} panel:} contrast as a function of the power of velocity oscillations. {\it Bottom middle panel:} contrast as a function of intensity oscillations, in arbitrary units. {\it Bottom right panel:} bi-dimensional representation of the \CaIIH\ contrast as a function of velocity and intensity oscillation power.} \label{fig:caii_wave}
\end{figure*}
%%%%%%%%%%%%%%%%%%%%%%%%%%%%%%%%%%%%%%%%%%%%%%%%%%%%%%%%%%%%%%

%%%%%%%%%%%%%%%%%%%%%%%%%%%%%%%%%%%%%%%%%%%%%%%%%%%%%%%%%%%%%%
\begin{figure*}
\centering
\includegraphics[width=12cm]{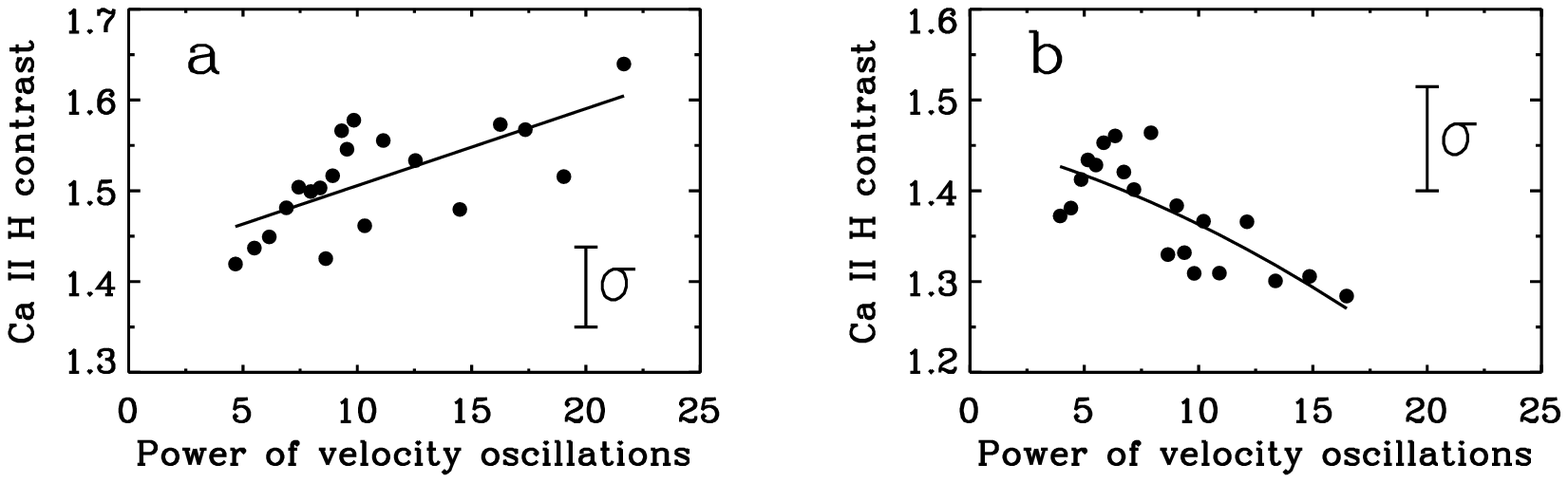}
\includegraphics[width=5cm]{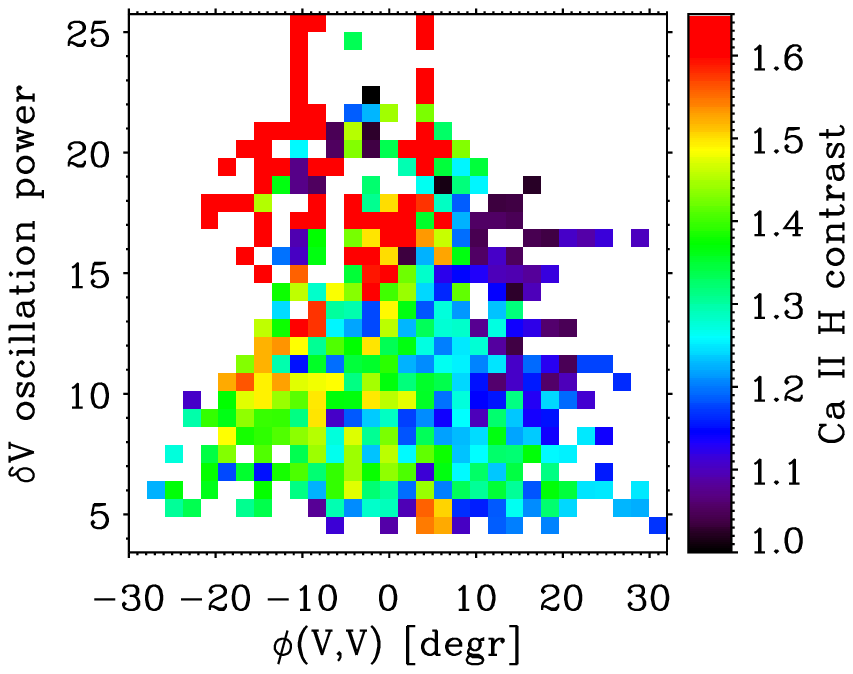}
\caption{{\it Left panel:} \CaIIH\ contrast as a function of power of upward propagating waves. {\it Middle panel:} contrast as a function of the power of downward propagating waves. {\it Right panel:} bi-dimensional representation of the \CaIIH\ contrast as a function of velocity phase shift, $\phi(V,V)$ indicative of the direction of the wave propagation, and velocity oscillation power. }\label{fig:waves}
\end{figure*}
%%%%%%%%%%%%%%%%%%%%%%%%%%%%%%%%%%%%%%%%%%%%%%%%%%%%%%%%%%%%%%

%%%%%%%%%%%%%%%%%%%%%%%%%%%%%%%%%%%%%%%%%%%%%%%%%%%%%%%%%%%%%%
\begin{figure*}
\centering
\includegraphics[width=12cm]{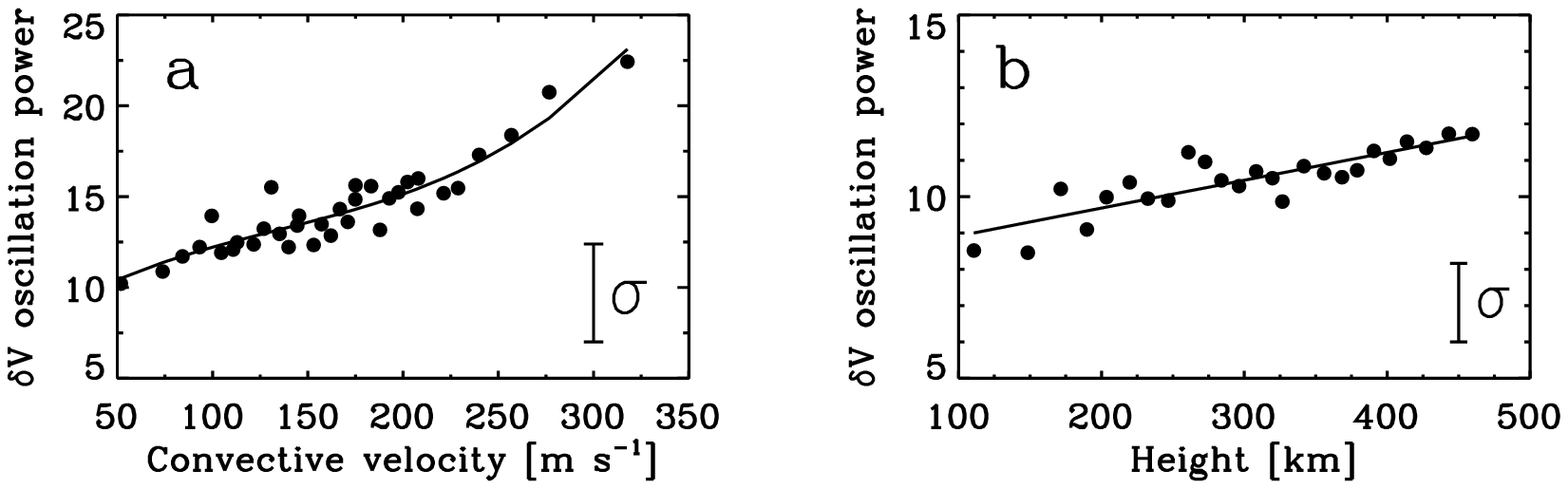}
\includegraphics[width=5cm]{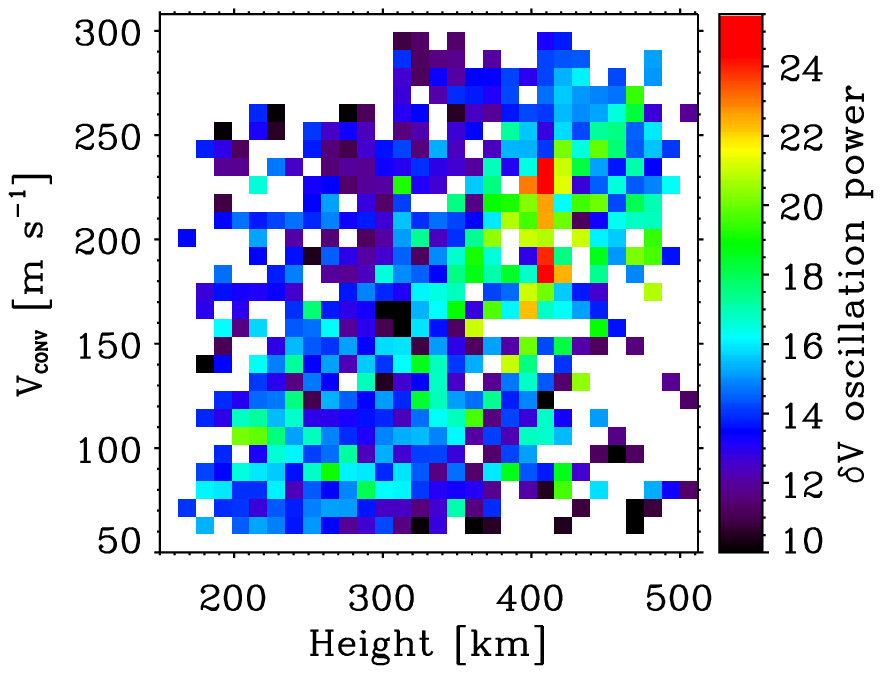}
\caption{{\it Left panel:} power of velocity oscillations in the \BaII\ line core as a function of the amplitude of convective motions at the photospheric base. {\it Middle panel:} same quantity as a function of height of sign reversal of convective motions. {\it Right panel:}  bi-dimensional representation of the velocity oscillation power as a function of the strength of convective motions and their reversal height. } \label{fig:wave-conv}
\end{figure*}
%%%%%%%%%%%%%%%%%%%%%%%%%%%%%%%%%%%%%%%%%%%%%%%%%%%%%%%%%%%%%%

\section{Results of observations}

Through the paper we use a statistical approach and search for correlations between different quantities. We use either one-dimensional dependences, or bi-dimensional ones. Figures \ref{fig:caii_mf_2d}--\ref{fig:caii_wave} show the dependence between the \CaIIH\ contrast and various parameters of convective and wave motions, and the magnetic field. In the case of one-dimensional dependences, each point represents the average of \CaIIH\ contrast in a bin with an equal number of data points in abscissa axis. The value of $\sigma$ shown in each panel is calculated as a average standard deviation between the individual and the average values in each bin, and then averaged for all bins. In the case of bi-dimensional plots, the \CaIIH\ contrast is averaged for the bins defined jointly for two independent quantities. The latter representation allows to study dependences between three parameters simultaneously.

Figure \ref{fig:caii_mf_2d} gives a bi-dimensional dependence between the contrast in \CaIIH\ and the photospheric magnetic field strength and inclination. We find a clear trend of increasing the contrast from 1.3 to 1.6 with magnetic field strength increasing from 350 to 1400 G, with a slight plateau leading to a slightly decreasing contrast around 1650 G. The dependence of the contrast on inclination is more monotonic: the contrast decreases from 1.65 to 1.3 with the inclination increasing from 5\degree\ to 40\degree. As can be seen in the figure, these two trends are not independent. Only a particular combination of the field strengths and inclinations is found in our data, in a way that stronger fields are more vertical. At the same time, for those stronger and vertical fields the values of \CaIIH\ contrast exceed those for weaker and more horizontal fields, as is indicated by the color scheme of the Figure \ref{fig:caii_mf_2d}. This trend is also seen in the maps in Figure \ref{fig:maps}. 

The contrast of the facula as a function of the parameters of convective motions is shown in Figure \ref{fig:caii_conv}. We obtain almost a linear dependence between the \CaIIH\ contrast and the magnitude of the convective velocity in the photosphere (left), and also between the contrast and the height where convective motions change their sign (middle).  In both cases the contrast increases with increasing amplitude of the convective motions and the height where the velocity sign reversal occurs. In our early works we have shown that not only intensity contrast reversal occurs in granulation, but also the reversal of the velocity sign  \citep{Kostik+Khomenko2012}. The plasma was shown to overshoot to higher heights at locations with stronger convective flows at the bottom photosphere, and also at locations with larger magnetic field. The right panel of Figure \ref{fig:caii_conv} completes this picture demonstrating that at locations of higher convective velocities and higher overshooting height the \CaIIH\ contrast of a plage region also increases. 

Figure \ref{fig:caii_wave} demonstrates how the wave motions influence the \CaIIH\ contrast. The left upper panel (marked ``a'') reveals that the contrast gets larger at locations where the phase shift $\phi(V,V)$ between velocities at the bottom and upper photosphere (as determined from \BaII\ profiles) is negative. According to our notation, a negative velocity phase shift means that waves propagate upwards. The phase shift is measured independently at each spatial point, at a frequency with a maximum cross-correlation coefficient between oscillations at both heights. This frequency lies in all cases in the five-minute band, with periods that vary between 270 and 330 sec. Such variations in periods does not influence the results presented in Fig. \ref{fig:caii_wave} since there is only a weak  dependence between the period and the $\phi(V,V)$ shift. This dependence is such that waves with larger periods have slightly larger negative $\phi(V,V)$ shift, while for waves with smaller periods  $\phi(V,V)$ shift turns positive. The 3 mHz waves observed in the photosphere are essentially evanescent, therefore their velocity phase shift with height is expected to be close to zero from theoretical considerations. Small non-zero phase shifts can be due to 5-min wave propagation allowed either by the ramp effect of inclined magnetic field lines, or by radiative losses. Here we find that at locations with waves traveling upwards (i.e. larger negative velocity phase shifts) the brightness of the facula gets larger. The brightness increases with increasing upward propagation speed (larger negative values of the velocity phase shift). 

The dependence of the \CaIIH\ contrast on the temperature-velocity phase shift, $\phi(T,V)$, of upper photospheric oscillations measured from the line core variations of \BaII\ has a more complex shape (upper middle panel in Fig. \ref{fig:caii_wave} marked ``b''). To get the temperature velocity phase shift we measured the velocity-intensity phase shift for oscillations with maximum power (periods in the range 270-330 sec in the photosphere) and assumed that temperature and intensity oscillations are 180 degrees out of phase. This conversion was done because we deal with a line of a singly ionized element. An increase in temperature leads to an increase in the number of atoms in the ionized state absorbing the radiation. The line becomes deeper and its intensity decreases; i.e.,\ the temperature and intensity oscillations are 180 degrees out of phase, see also \citet{Shchukina+etal2009}.  Similar assumption was done in \citet{Kostik+Khomenko2013}. We operate in terms of temperature-velocity phase shift for a better comparison with theoretical models, see \citet{Noyes+Leighton1963, Mihalas+Toomre1981,Mihalas+Toomre1982, Deubner1990}. The facular contrast is minimum for phase shift values around $\pm180$\degree\ and has a broad maximum between 0\degree\ and 100\degree.

The right upper panel of Figure \ref{fig:caii_wave} completes the picture of the contrast dependence on the wave phase shifts by showing a bi-dimensional dependence. It reveals that not all combinations of the $\phi(T,V)$ and $\phi(V,V)$ phase shifts are present in our data.  Locations with enhanced \CaIIH\ contrast correspond to the locations where two conditions are simultaneously satisfied: the $\phi(V,V)$ phase shift is essentially negative, i.e. waves are propagating upwards, and the $\phi(T,V)$ phase shift is comprised between 0\degree\ and 120\degree. Such values of the temperature-velocity phase shift are characteristic for acoustic-gravity waves strongly affected by radiative losses, and have been frequently detected in the earlier observations, see \citet{Noyes+Leighton1963, Deubner+Fleck1989, Fleck+Deubner1989}. Other combinations of the $\phi(T,V)$ and $\phi(V,V)$ phase shifts are also present. For positive values of $\phi(V,V)$, i.e. downward wave propagation, the $\phi(T,V)$ shifts are essentially either negative or above 120\degree. In those cases, the \CaIIH\ contrast is at its lowest values. 

The lower panels of Figure \ref{fig:caii_wave} give the dependence between the power of velocity and intensity oscillations in the upper photosphere, and the \CaIIH\ contrast. There is a weak evidence that the contrast increases with increasing velocity oscillation power, while it is independent of the intensity oscillation power. The bi-dimensional representation of the same dependences provided at the lower right panel confirms this conclusion: larger \CaIIH\ contrast is observed in the chromsphere at locations where the velocity oscillations are stronger, but there is a large scatter for the intensity oscillations. Similar conclusions can be also derived by the visual comparison of the spatial distribution of the velocity oscillation power and the power of \CaIIH\ intensity oscillations from the middle panels of Figure \ref{fig:maps}. The patches with enhanced intensity oscillations are smaller and more localized and the dependence between the contrast the the intensity oscillation power is not apparent.

%%%%%%%%%%%%%%%%%%%%%%%%%%%%%%%%%%%%%%%%%%%%%%%%%%%%%%%%%%%%%%
\section{Discussion}
%%%%%%%%%%%%%%%%%%%%%%%%%%%%%%%%%%%%%%%%%%%%%%%%%%%%%%%%%%%%%%

The dependences between the chromospheric facular brightness and the parameters of oscillations (Fig. \ref{fig:caii_wave}) are in agreement with an intuitive picture of upward running waves transferring their energy to the chromosphere and producing an increase of its brightness. Indeed, according to the analysis of the same dataset in \citep{Kostik+Khomenko2013}, at about 67\% of locations in the observed area the $\phi(V,V)$ is negative, i.e. in most of the area the waves are propagating upwards. 
The larger is the power of velocity oscillations, the larger is the facular contrast (Fig. \ref{fig:caii_wave}c). Intensity oscillations affect less the facular contrast, possibly because of the magnetic nature of the observed waves, in which case the principal restoring force is magnetic field and oscillations of thermodynamic parameters are less important.

Figure \ref{fig:waves} provides additional details on the relation between the wave propagation direction, their power and \CaIIH\ contrast. The left and middle panels give similar dependences as the panel (c) of Fig. \ref{fig:caii_wave} but this time separating upward and downward running waves. We obtain that the contrast increases with the wave power only for upward running waves (left panel). For the downward running ones (right panel), the dependence is just the opposite. Similar conclusion is also apparent from the bi-dimensional representation of the contrast as a function of the velocity oscillation power and $\phi(V,V)$, see right panel of Figure \ref{fig:waves}. The contrast is maximum where two conditions are simultaneously satisfied: waves are propagating upwards and their power is maximum. This result needs further studies. 

As was shown in our previous work \citep{Kostik+Khomenko2013} the 5 minute oscillations reach more effectively the temperature minimum from the bottom photosphere when the phase shift between the oscillations of temperature and velocity are in the range of 0\degree\ and 90\degree. If such waves were observed in a quiet area, such phase shift would be suggestive for clear deviations from adiabatic wave propagation (an adiabatic propagation leads to 180\degree\ phase shift). While $\phi(V,V)$ phase shifts allow for a more straightforward interpretation in terms of the direction of the wave propagation, it is not so for the $\phi(T,V)$ phase shifts \citep{Deubner1990}. Solar 5-minute oscillations are strongly affected by radiative losses and the precise value of the $\phi(T,V)$ phase shift depends of the amount of radiative losses, wave frequency and the height in the atmosphere where the waves are observed \citep[see][among others]{Mihalas+Toomre1981, Mihalas+Toomre1982, Deubner1990, Deubner+Fleck1989, Fleck+Deubner1989}.  Due to the rapid change with height of the sound speed (temperature), wave reflection can also occur and standing waves can be produced. In such a case the values of the $\phi(T,V)$ phase shift of 90\degree\ would also appear if the wave propagation is considered adiabatic. However, there are arguments against such interpretation. On the one hand,  there have been theoretical studies of the acoustic-gravity wave reflection \citep[e.g.,][]{Marmolino+Severino+Deubner+Fleck1993} which have shown that in the evanescent region of the $k-\omega$ diagram between the Lamb frequency and the acoustic cut-off frequency (i.e. in the range of frequencies similar to that in our data), the reflection coefficient is small. On the other hand, standing waves can be discarded by the upper right panel of Figure \ref{fig:caii_wave} showing negative values of the $\phi(V,V)$ phase shifts (upward propagation) when $\phi(T,V)$ phase shifts lie in the range between  0\degree\ and 90\degree. One, however, has to take additional care in interpreting $\phi(T,V)$ phase shifts in our case, since the waves are observed in a strongly magnetized facular area and the dependence between temperature-velocity phase shift and non-adiabaticity of oscillations in the presence of magnetic field is not well studied theoretically \citep[see the discussion in][]{Kostik+Khomenko2013}. Determining the phase shift between temperature and velocity for magneto-acoustic waves would need numerical modeling for particular cases of the observed magnetic field geometries. 

It is intriguing that the facular contrast in the chromosphere depends on the strength of convective motions in the photosphere, given that flows do not reach chromospheric heights. \citet{Kostik+etal2006} finds that the power of 5-minute oscillations at the formation height of the \FeI\ 6393.6 \AA\ spectral line (around 500 km) increases with increasing strength of convective motions at the bottom photosphere. It is then possible that something similar happens for the \BaII\ line. Figure \ref{fig:wave-conv} provides the dependence between the power of 5-minute oscillations at the core of \BaII\ line and the convective velocity, demonstrating that the wave power increases with increasing strength of convective motions, similar to what was found in \citet{Kostik+etal2006}. The wave power also increases with increasing height where the convective motions change their sign (middle panel of Fig. \ref{fig:wave-conv}), and is maximum at locations where both conditions (large convective velocity and higher reversal height) are satisfied, see right panels of Fig. \ref{fig:wave-conv}. Therefore, it can be concluded that convective motions, by exciting oscillations, indirectly contribute to the increase of facular brightness in the chromosphere. 

%%%%%%%%%%%%%%%%%%%%%%%%%%%%%%%%%%%%%%%%%%%%%%%%%%%%%%%%%%%%%%
\section{Conclusions}
%%%%%%%%%%%%%%%%%%%%%%%%%%%%%%%%%%%%%%%%%%%%%%%%%%%%%%%%%%%%%%

Due to the presence of magnetic field in the facular area, 5-minute waves penetrate to chromospheric heights (either along the magnetic field lines or due to non-adiabatic effects), and lead to an efficient energy transfer to the chromosphere of solar facula. The brightness of the facula was found to strongly depend on the power of waves traveling upwards. The convective motions at the photospheric base influence the brightness in the indirect way. The larger is the amplitude of convective motions and the height in the photosphere where they change sign, the brighter is the facula. All these results together lead to the conclusion that facular areas appear bright not only due to the Wilson depression, but also because of a real heating.

\begin{acknowledgements}
This work is partially supported by the Spanish Ministry of Science through projects AYA2010-18029, AYA2011-24808 and AYA2014-55078-P. This work contributes to the deliverables identified in FP7 European Research Council grant agreement 277829, ``Magnetic connectivity through the Solar Partially Ionized Atmosphere''. 
\end{acknowledgements}

%\aareferences

\end{document}